\documentclass[aps,pre,groupedaddress,superscriptaddress,nofootinbib,twocolumn]{revtex4}

\usepackage{graphicx,amsmath,amssymb,amsbsy,subfigure,hyperref,bbm,times,txfonts,float}
\usepackage{subfigure,hyperref,bbm,times}
\usepackage[T1]{fontenc}
\usepackage{braket}
\usepackage{epsfig} 
\usepackage{color}
\usepackage{amsmath}
\usepackage{xcolor}%
\usepackage{graphicx}
\usepackage{dcolumn}
\usepackage{bm}

\providecommand{\openone}{\leavevmode\hbox{\small1\kern-3.8pt\normalsize1}}

\usepackage{soul}

\hypersetup{
	colorlinks=true,
	linkcolor=blue,
}
\graphicspath{{figure/}}

\begin{document}

	\title{Collective-dissipation-controlled thermal rectification and entropy production in a three-terminal three-qubit XXZ spin chain}
	\author{Alireza Nourmandipour}
	\email{anourmandip@sirjantech.ac.ir}
	\affiliation{Department of Physics, Sirjan University of Technology, 7813733385 Sirjan, Iran}
	
\begin{abstract}
	We investigate thermal rectification and its thermodynamic cost in a three-qubit XXZ spin chain coupled to three thermal reservoirs: two local baths attached to the boundary qubits and a collective bath jointly coupled to the middle and right qubits. This asymmetric dissipative architecture allows the collective environment to actively reshape energy transport. We show that the heat current carried by the collective bath changes sign as its temperature is varied, and that the spin-chain anisotropy provides an effective internal control parameter for suppressing the boundary heat currents. These features yield thermal rectification coefficients of about $3.36\%$ at maximum bias and up to $20.1\%$ in the strongly anisotropic regime, although the latter comes at the cost of severely reduced heat currents. Beyond rectification, we evaluate the steady-state entropy production rate and identify a low-irreversibility operating window near the collective-current reversal, whereas increasing the local boundary couplings monotonically raises the thermodynamic cost. Our results establish collective dissipation as a tunable control mechanism for quantum heat transport and uncover a nontrivial trade-off among rectification, heat-current magnitude, and irreversible losses.
\end{abstract}
	
	\date{\today }
	
	\maketitle
	
\section{Introduction}

Open quantum systems have become a central paradigm in modern physics, providing the theoretical framework for understanding how microscopic systems interact with their environments \cite{Diehl2008,Kraus2008,Müller2012,Daley2014,Ikeda2020}. While much of the early work in this area focused on decoherence and dissipation as obstacles to quantum control \cite{Rafiee2020,Chen2021}, a growing body of research has revealed that environment-induced effects can be harnessed as resources. The interplay between coherent dynamics and dissipation gives rise to rich phenomena such as quantum synchronization \cite{Almani2026}, dissipative phase transitions \cite{Matsumoto2025}, and entanglement generation through reservoir engineering \cite{Vafafard2022}. In the context of quantum thermodynamics, this perspective has opened up new avenues for designing thermal devices that exploit, rather than merely tolerate, dissipative effects \cite{alicki1979}. 

The manipulation of heat flow at the nanoscale has emerged as a central theme in modern quantum thermodynamics, driven by both fundamental questions about the nature of non-equilibrium steady states and the practical need for efficient thermal management in quantum devices \cite{gemmer2009, kosloff2013}. Among the most sought-after functionalities is the quantum thermal diode—a device that allows heat to flow preferentially in one direction while suppressing it in the reverse \cite{li2004, segal2006}. Early theoretical proposals demonstrated that thermal rectification can arise from asymmetric couplings, nonlinear interactions, or broken spatial symmetries in low-dimensional quantum systems \cite{chang2006, werlang2014}. Recent advances in circuit quantum electrodynamics and trapped-ion platforms have brought these ideas closer to experimental realization, where quantum spin chains serve as ideal testbeds for exploring energy transport under controlled conditions \cite{you2011, blatt2012}.

Despite the progress achieved in quantum thermal rectification, an important question remains: can the dissipative environment itself be exploited as an active resource for controlling heat transport? In conventional two-terminal configurations, rectification is most often associated with asymmetries built into the system, such as spatially dependent couplings, onsite energies, or other Hamiltonian properties \cite{Pereira2019}. In contrast, recent studies of nonequilibrium open quantum systems have shown that appropriately engineered dissipation can itself generate directional particle transport \cite{yamamoto2020}. This perspective suggests moving beyond static structural asymmetry toward reservoir engineering, in which an additional environmental degree of freedom provides an externally tunable means of modifying the transport pathways. In particular, a collective reservoir coupled simultaneously to multiple sites can introduce correlated dissipative processes and nonlocal energy-exchange pathways, offering a route to modify heat transport \cite{Huang2022}.

In many two-terminal quantum thermal diodes, rectification is generated by asymmetries encoded in the system itself. Segmented XXZ chains provide a clear realization of this mechanism, where strong anisotropic interactions in one part of the chain can suppress heat transport under reverse bias and thereby produce sizeable rectification \cite{balachandran2018,werlang2014}. Such Hamiltonian-engineering strategies demonstrate that strong directionality can emerge from the internal structure of the device. However, they also motivate a different question: can the preferred direction of heat transport instead be controlled through the dissipative environment, and at what thermodynamic cost?

In this work, we address this problem using a minimal three-qubit XXZ spin chain coupled to three thermal terminals. The two boundary qubits are connected to independent left and right reservoirs, while the two central qubits are jointly coupled to a common collective reservoir through the symmetric jump operator $\Sigma=\sigma_2^-+\sigma_3^-$. The resulting three-terminal dissipative architecture breaks the left-right symmetry of the boundary environments while retaining a symmetric collective coupling between the central qubits. This allows the collective reservoir to serve as an externally tunable dissipative control element, providing a means of modifying both the magnitude and direction of steady-state heat currents. Using a Lindblad master equation with thermal dissipative channels, we systematically investigate how collective dissipation and the interaction anisotropy shape heat transport and thermal rectification.

Our results reveal several distinct physical features. The collective reservoir can undergo a crossover from absorbing energy from the system to supplying energy to it as its thermodynamic state is varied, leading to a reversal of the associated heat current. We further find that the interaction anisotropy strongly suppresses the boundary heat currents and can enhance thermal rectification, although stronger rectification is accompanied by substantially reduced heat-current magnitudes. Because rectification alone does not characterize the thermodynamic cost of operation, we additionally evaluate the steady-state entropy production and identify regimes of particularly low irreversibility. We also find that stronger coupling to the local boundary reservoirs systematically increases entropy production. Taken together, these results demonstrate that collective dissipation can serve as an active resource for controlling quantum heat transport and reveal a nontrivial trade-off among transport magnitude, rectification, and thermodynamic irreversibility.

The remainder of this paper is organized as follows. Section \ref{sec:Model} introduces the three-qubit XXZ model. Section \ref{sec:heat_transport} presents the steady-state heat currents. Section \ref{sec:rectification} investigates thermal rectification and its dependence on the system parameters. Section \ref{sec:entropy_production} examines the associated entropy production and discusses the thermodynamic cost of directional heat transport. Finally, Section \ref{sec:conclusion} summarizes the main results.

\section{Model}
\label{sec:Model}
	
We consider an open quantum system consisting of three spin-\(\frac{1}{2}\) particles (qubits) arranged in a linear chain and interacting through an anisotropic XXZ Heisenberg coupling. Such a minimal spin network provides a convenient platform for studying the interplay between quantum correlations, dissipation, and nonequilibrium heat transport in engineered quantum devices.
	
	The system Hamiltonian is given by
	\begin{equation}
		H_S =
		\frac{1}{2}\sum_{i=1}^{2}
		\left[
		J\left(\sigma_i^x \sigma_{i+1}^x + \sigma_i^y \sigma_{i+1}^y\right)
		+ \Delta \sigma_i^z \sigma_{i+1}^z
		\right]
		+
		\sum_{i=1}^{3}\frac{B_0}{2}\sigma_i^z,
	\end{equation}
	where $\sigma_i^\alpha$ ($\alpha=x,y,z$) denote the Pauli operators acting on qubit $i$, $J>0$ is the exchange coupling strength, $\Delta$ is the anisotropy parameter in the $z$ direction, and $B_0$ is a uniform external magnetic field. This model reduces to the XX limit for $\Delta=0$, to the isotropic Heisenberg case for $\Delta=J$ in the present parametrization, and to an Ising-type interaction for $J=0$.
	\begin{figure}[htbp]
		\centering
		\includegraphics[width=0.45\textwidth]{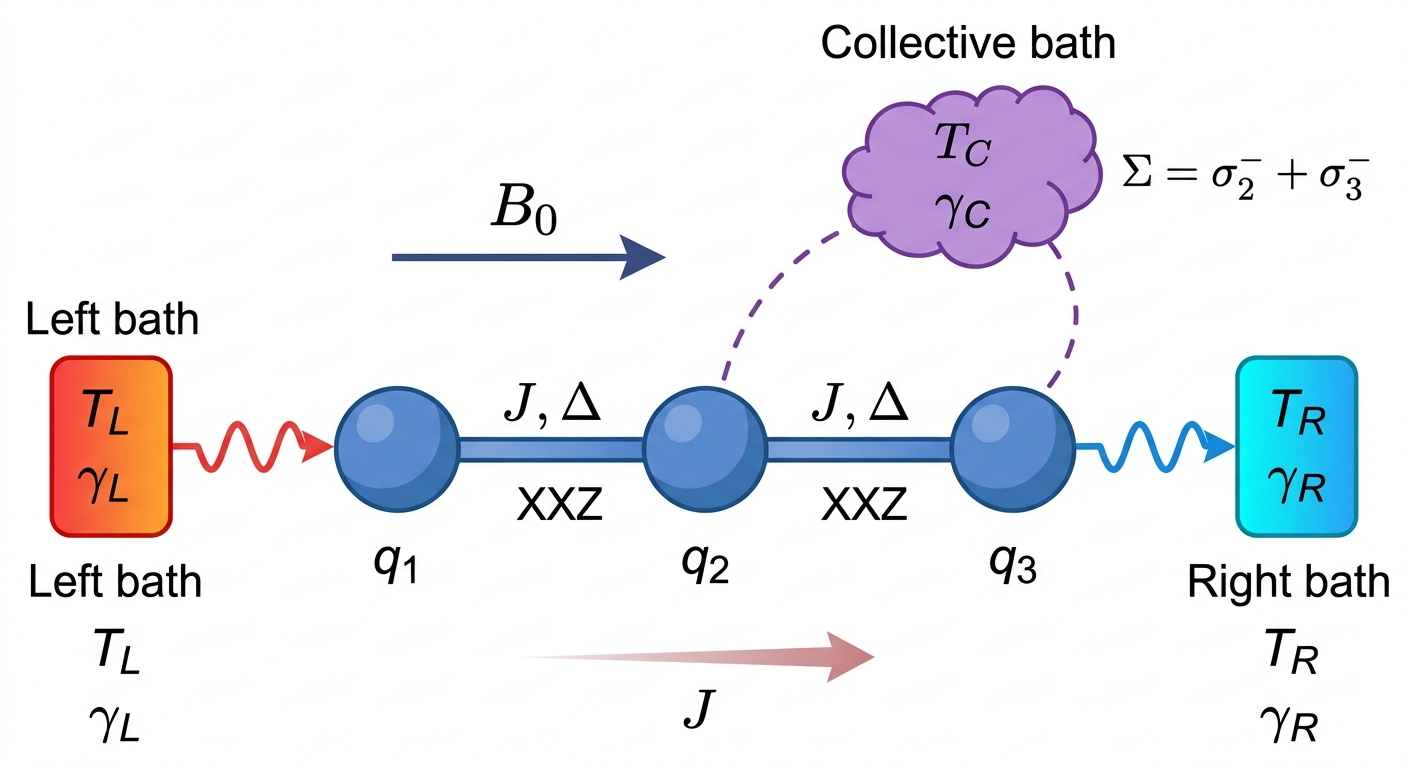}
		\caption{Three-qubit XXZ spin chain coupled to local and collective thermal reservoirs. Qubits 1 and 3 are attached to boundary baths at temperatures $T_L$ and $T_R$, respectively, while qubits 2 and 3 are jointly coupled to a common bath at temperature $T_C$. The asymmetric collective dissipation constitutes the nonequilibrium control mechanism responsible for modifying the steady-state heat currents and enabling thermal rectification.}
		\label{Fig_1}
	\end{figure}
	To drive the system out of equilibrium, qubit $1$ is coupled to a left thermal reservoir at temperature $T_L$, while qubit $3$ is coupled to a right thermal reservoir at temperature $T_R$. In addition, qubits $2$ and $3$ interact collectively with a common reservoir at temperature $T_C$. This collective environment is introduced through a symmetric jump operator and acts as an additional dissipative channel that can modify the steady-state energy transport through correlated decay processes. Because the collective reservoir couples jointly to qubits $2$ and $3$, while the boundary qubits are coupled to independent reservoirs, the resulting three-terminal dissipative architecture is asymmetric with respect to the two boundaries and can therefore modify the directional heat transport.
	
	The reduced density matrix $\rho(t)$ of the system evolves according to the Lindblad master equation \cite{Lindblad1976,Gorini1976}
	\begin{equation}
		\frac{d\rho}{dt}
		=
		-i[H_S,\rho]
		+
		\mathcal{D}_L[\rho]
		+
		\mathcal{D}_R[\rho]
		+
		\mathcal{D}_C[\rho],
	\end{equation}
	where $\mathcal{D}_L[\rho]$, $\mathcal{D}_R[\rho]$, and $\mathcal{D}_C[\rho]$ describe the dissipative action of the left, right, and collective reservoirs, respectively. The local thermal dissipators acting on the boundary qubits are taken in the phenomenological Lindblad form
	\begin{equation}
		\mathcal{D}_L[\rho]
		=
		\gamma_L (n_L+1)\,\mathcal{L}[\sigma_1^-]\rho
		+
		\gamma_L n_L\,\mathcal{L}[\sigma_1^+]\rho,
	\end{equation}
	\begin{equation}
		\mathcal{D}_R[\rho]
		=
		\gamma_R (n_R+1)\,\mathcal{L}[\sigma_3^-]\rho
		+
		\gamma_R n_R\,\mathcal{L}[\sigma_3^+]\rho,
	\end{equation}
	where $\sigma_i^\pm = (\sigma_i^x \pm i\sigma_i^y)/2$ are the spin ladder operators, $\gamma_L$ and $\gamma_R$ are the system-bath coupling strengths, while the collective dissipation acting on qubits $2$ and $3$ is modeled as
	\begin{equation}
		\mathcal{D}_C[\rho]
		=
		\gamma_C (n_C+1)\,\mathcal{L}[\Sigma]\rho
		+
		\gamma_C n_C\,\mathcal{L}[\Sigma^\dagger]\rho,
	\end{equation}
	with the collective lowering operator
	\begin{equation}
		\Sigma = \sigma_2^- + \sigma_3^-,
	\end{equation}
	where $\gamma_C$ denotes the collective dissipation strength and $n_C$ is the thermal occupation number of the common reservoir evaluated at the characteristic transition frequency $\omega$. The use of the operator $\Sigma$ introduces cross terms between the dissipative channels of qubits $2$ and $3$, allowing the collective reservoir to modify the effective transport pathways and redistribute energy flow among the three terminals. The Lindblad superoperator is defined as $\mathcal{L}[A]\rho = A\rho A^\dagger - \frac{1}{2}\left\{A^\dagger A,\rho\right\}.$
	
	In the above relations
	\begin{equation}
		n_\nu = \frac{1}{e^{\omega/T_\nu}-1},
		\qquad \nu=L,R,C,
	\end{equation}
	is the Bose-Einstein occupation number evaluated at a characteristic frequency $\omega$, which is taken as $\omega=1$ for simplicity.

	The present setup should therefore be regarded as a three-terminal open quantum system, in which the collective reservoir acts as an environmental control channel for the transport between the left and right baths.

	\section{Heat Transport}
	\label{sec:heat_transport}
	
	Having established the system Hamiltonian and dissipative dynamics, we now turn to the steady-state energy flow through the three-qubit chain. The heat currents associated with the local and collective reservoirs constitute the primary transport observables of the present work, as they directly characterize how energy is exchanged among the baths and how collective dissipation modifies transport through the chain. In this section, we focus on the steady-state heat currents themselves---their magnitudes, their dependence on the thermal bias, and their sensitivity to the collective-bath parameters---before addressing thermal rectification in Sec.~\ref{sec:rectification}.
	
	To establish the steady-state energy balance, we note that 	the heat current associated with reservoir $\alpha$ ($\alpha=L,R,C$) is defined at the nonequilibrium steady state $\rho_{\mathrm{ss}}$ as
	\begin{equation}
		J_\alpha
		=
		\mathrm{Tr}\left[ H_S\,\mathcal{D}_\alpha(\rho_{\mathrm{ss}}) \right],
	\end{equation}
	where a positive value of $J_\alpha$ corresponds to energy flowing from reservoir $\alpha$ into the system. 
	The system energy evolves according to
	\begin{equation}
		\frac{d}{dt}\langle H_S\rangle = \mathrm{Tr}\left(H_S\dot{\rho}\right).
	\end{equation}
	Using the master equation in Eq.~(2), we obtain
	\begin{equation}
		\frac{d}{dt}\langle H_S\rangle
		=
		-i\,\mathrm{Tr}\left\{H_S[H_S,\rho]\right\}
		+
		\sum_{\alpha=L,R,C}
		\mathrm{Tr}\left\{H_S\mathcal{D}_\alpha(\rho)\right\}.
	\end{equation}
	The coherent (unitary) contribution vanishes identically by virtue of the cyclic property of the trace,
	\begin{equation}
		\mathrm{Tr}\left\{H_S[H_S,\rho]\right\}
		=
		\mathrm{Tr}\left\{[H_S,H_S]\rho\right\}
		=0,
	\end{equation}
	where we used $\mathrm{Tr}(A[B,C])=\mathrm{Tr}([A,B]C)$. Recognizing the dissipative contributions as the heat currents defined above, we have
	\begin{equation}
		\frac{d}{dt}\langle H_S\rangle = J_L+J_R+J_C.
	\end{equation}
	At the nonequilibrium steady state, the internal energy of the system is stationary, $d\langle H_S\rangle/dt=0$, which immediately yields the energy-conservation condition
	\begin{equation}
		\boxed{
			J_L+J_R+J_C=0.
		}
	\end{equation}

	We then examine how the boundary and collective heat currents depend on the relevant control parameters, including the thermal bias, the collective coupling strength $\gamma_C$, the collective bath temperature $T_C$, and, where appropriate, the anisotropy parameter $\Delta$. This allows us to identify the transport regimes in which the collective environment acts as an additional energy-exchange channel and significantly reshapes the effective thermal response of the device. These results provide the basis for the rectification analysis presented in the following section.
	
	\begin{figure}[htbp]
		\centering
		\includegraphics[width=0.45\textwidth]{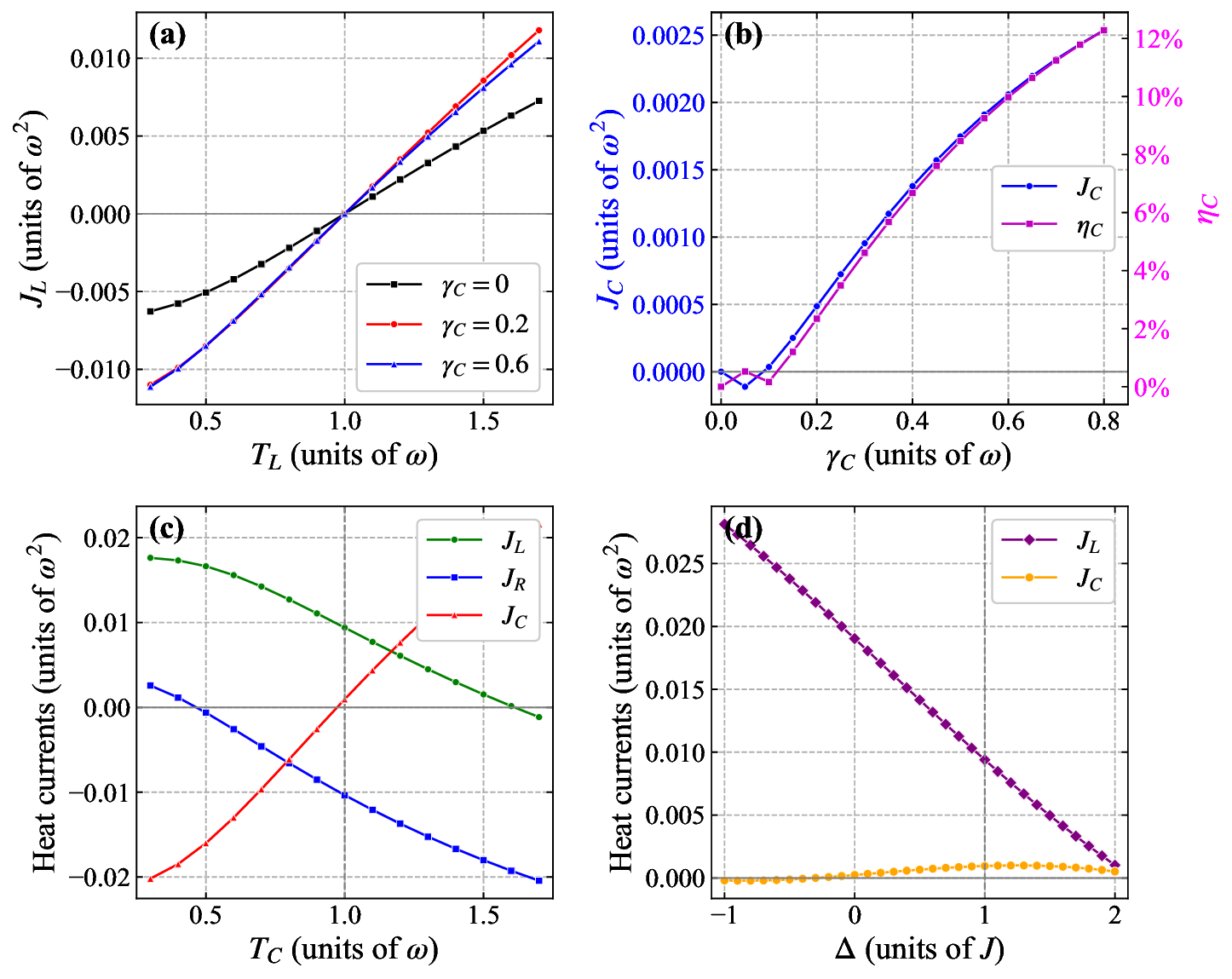}
		\caption{
			Steady-state heat currents in the three-qubit XXZ spin chain with local and collective dissipation. 
			(a) $J_L$ versus $T_L$ for different collective couplings $\gamma_C$, with $\gamma_L=\gamma_R=0.1$, $T_R=1$, $T_C=1$, $\Delta=1$. 
			(b) Collective current $J_C$ (left axis) and collective-channel weight $\eta_C$ (right axis) as functions of $\gamma_C$, with $\gamma_L=\gamma_R=0.1$, $T_L=1.5$, $T_R=0.5$, $T_C=1$, $\Delta=1$. 
			(c) $J_L$, $J_R$, and $J_C$ versus the collective-bath temperature $T_C$, with $\gamma_L=\gamma_R=0.1$, $T_L=1.5$, $T_R=0.5$, $\gamma_C=0.3$, $\Delta=1$. 
			(d) Boundary and collective currents as functions of the anisotropy parameter $\Delta$, with $\gamma_L=\gamma_R=0.1$, $T_L=1.5$, $T_R=0.5$, $T_C=1$, $\gamma_C=0.3$. 
			For all the plots $B_0=1$ and all currents are in units of $\omega^2$.
		}
		\label{Fig_2}
	\end{figure}
	
	Figure~\ref{Fig_2} presents the steady-state heat currents for several representative parameter scans and illustrates how the collective reservoir reshapes the transport pattern of the chain. We begin in Fig.~\ref{Fig_2}(a) with the dependence of the currents on the left-bath temperature $T_L$ for different values of the collective coupling strength $\gamma_C$, with $T_R=1$ and $T_C=1$. In the absence of collective dissipation, $\gamma_C=0$, the system reduces to an effectively two-terminal configuration: the collective current vanishes, $J_C=0$, and the boundary currents satisfy $J_L=-J_R$. The sign change of the currents near the thermal symmetry point $T_L=1$ marks the crossover between opposite directions of net energy flow. Moderate collective coupling ($\gamma_C=0.2$) enhances $J_L$ compared to the purely local case, reaching $J_L \approx 0.0118$ at $T_L=1.7$, whereas stronger coupling ($\gamma_C=0.6$) yields a similar or slightly smaller value, indicating that the bath-induced transport enhancement is not monotonic in $\gamma_C$. This nonmonotonic behavior confirms that the collective reservoir does not merely contribute additional damping, but instead participates actively in the exchange of energy with the spin chain.
	
	The effect of increasing $\gamma_C$ is quantified more clearly in Fig.~\ref{Fig_2}(b), where we plot both the collective current $J_C$ and the normalized collective current
	\begin{equation}
		\eta_C=\frac{|J_C|}{|J_L|+|J_R|+|J_C|}.
	\end{equation}
	As the collective coupling grows, $\eta_C$ increases steadily, reaching approximately $12.3\%$ at $\gamma_C=0.8$. At the same time, the boundary currents are not enhanced uniformly; rather, the growth of the collective contribution is accompanied by a redistribution of transport among the available channels. Specifically, the left boundary current $J_L$ becomes suppressed at larger $\gamma_C$ (decreasing by approximately $27.6\%$), while $J_R$ remains relatively stable. The main role of collective dissipation is therefore to reconfigure the current pathways, not simply to amplify the total current.
	
	This interpretation is reinforced by Fig.~\ref{Fig_2}(c), which shows the dependence of $J_L$, $J_R$, and $J_C$ on the collective-bath temperature $T_C$ for fixed boundary temperatures $T_L=1.5$ and $T_R=0.5$, with $\gamma_C=0.3$. Varying $T_C$ strongly modifies all three currents, confirming that the collective reservoir acts as an active thermodynamic control parameter. Most notably, the collective current changes sign near $T_C \approx 1$, indicating a crossover in the direction of heat flow through the collective channel. This sign reversal occurs when the collective bath temperature crosses a value determined by the balance between the hot and cold reservoirs. For $T_C$ below this effective temperature, the bath extracts heat from the system ($J_C < 0$); for $T_C$ above it, the bath injects heat ($J_C > 0$). According to our sign convention, this means that the collective bath may either inject energy into the chain or extract energy from it, depending on the global temperature configuration. For $T_C < 1$, the collective bath extracts heat from the system ($J_C < 0$); for $T_C > 1$, it injects heat ($J_C > 0$). It is therefore more accurate to regard the collective environment as a tunable third terminal whose role can change from source to sink under nonequilibrium conditions.
	
	The influence of the XXZ anisotropy parameter is displayed in Fig.~\ref{Fig_2}(d), where we fix $T_L=1.5$, $T_R=0.5$, $T_C=1$, and $\gamma_C=0.3$. As $\Delta$ increases from negative to positive values, the magnitudes of the boundary currents decrease markedly, with $J_L$ dropping from $J_L \approx 0.0281$ at $\Delta=-1$ to $J_L \approx 0.00102$ at $\Delta=2$, a reduction by a factor of approximately $28$. This suppression can be understood as a consequence of excitation localization: for positive $\Delta$, the Ising-like interaction $\Delta \sigma_i^z \sigma_{i+1}^z$ dominates over the $XY$ exchange, effectively freezing spin-flip processes that are responsible for energy transport. Conversely, for negative $\Delta$ (ferromagnetic regime), the system favors spin alignment, which enhances coherent propagation of excitations and thus increases the heat current. By contrast, the collective current $J_C$ remains smaller in magnitude and exhibits a weaker, non-monotonic dependence on $\Delta$, peaking near $\Delta \approx 1.2$. The dominant effect of anisotropy is thus the attenuation of the boundary transport channels, while the collective pathway remains secondary over most of the explored parameter range.
	
The results of Fig.~\ref{Fig_2} establish that collective dissipation opens an additional energy-exchange channel and breaks the simple two-terminal balance, that the collective reservoir can either absorb or supply heat depending on the thermal bias, and that the anisotropy parameter $\Delta$ provides an efficient mechanism for suppressing the overall transport response. Throughout all parameter scans, the numerical solutions satisfy the steady-state energy-balance condition to machine precision, and the corresponding density matrices remain positive semidefinite, confirming the physical consistency and numerical stability of the reported currents.

\section{Thermal Rectification}
\label{sec:rectification}

To evaluate the potential for thermal-diode behavior, we quantify the directional asymmetry of heat transport under reversal of the temperature bias. Fixing the collective-bath parameters (in particular $T_C = 1$), we compare the left-reservoir current in the forward configuration, $J_L^{\mathrm{fwd}}$, obtained with $T_L > T_R$, against the current in the reverse configuration, $J_L^{\mathrm{rev}}$, obtained by interchanging the reservoir temperatures ($T_L \leftrightarrow T_R$). A convenient rectification coefficient is defined as \cite{Karg2019}
\begin{equation}
	\mathcal{R} = \frac{|J_L^{\mathrm{fwd}}| - |J_L^{\mathrm{rev}}|}{\max\left(|J_L^{\mathrm{fwd}}|, |J_L^{\mathrm{rev}}|\right)},
\end{equation}
where $\mathcal{R}=0$ denotes perfectly reciprocal transport (no rectification), while $\mathcal{R}\neq 0$ signals a directional preference. With this definition, $\mathcal{R} \in [-1, 1]$, and a positive value indicates that heat flows more readily from left to right than from right to left. Because the system is three-terminal, the rectification protocol is defined by interchanging the temperatures of the two boundary reservoirs while keeping the collective-reservoir parameters fixed.
	
\begin{figure}[htbp]
		\centering
		\includegraphics[width=0.45\textwidth]{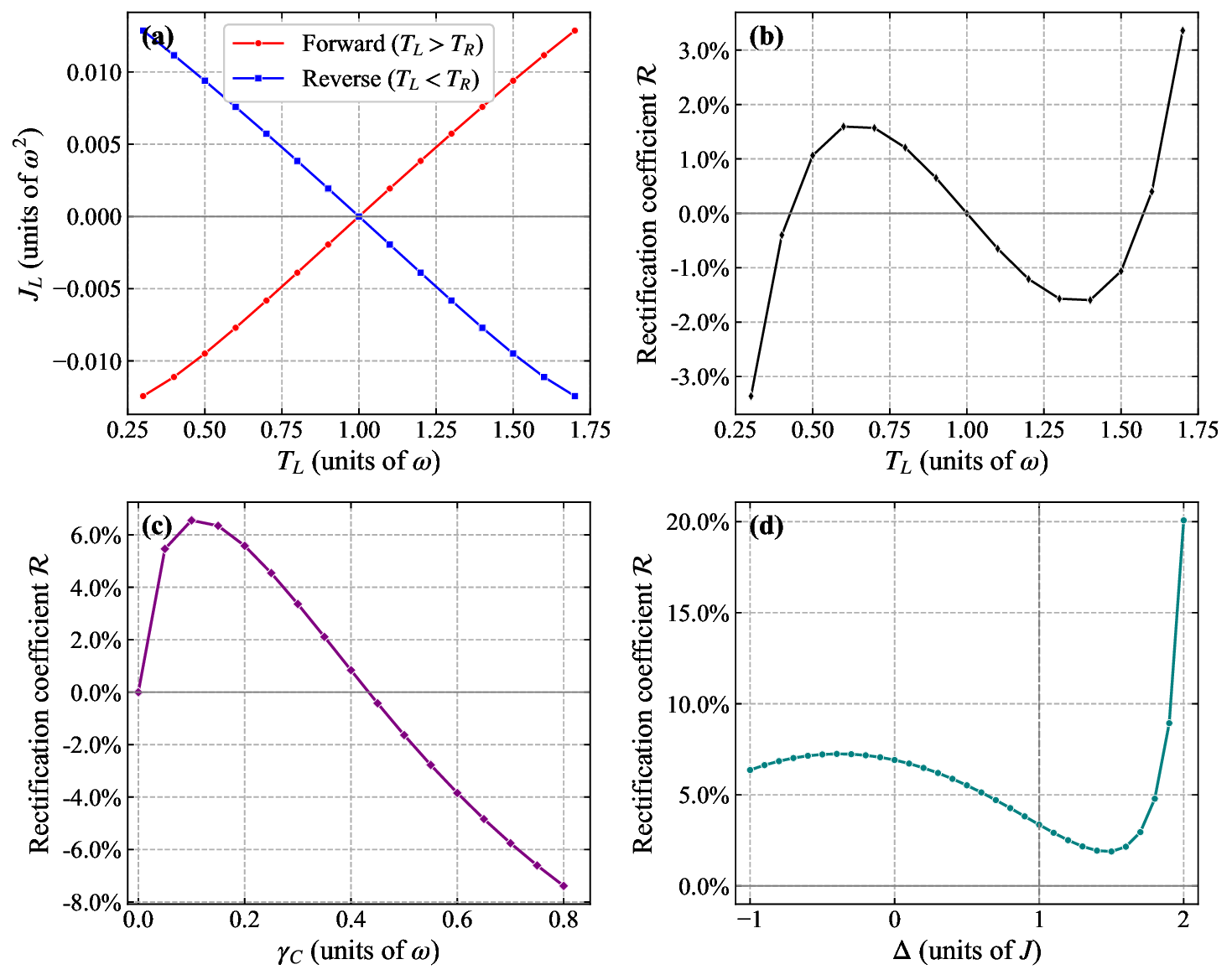}
		\caption{
			Thermal rectification in the three-qubit XXZ chain with collective dissipation. (a) Heat current $J_L$ as a function of $T_L$ for forward bias ($T_L > T_R$, red circles) and reverse bias ($T_L < T_R$, blue squares), under the constraint $T_L + T_R = 2$, with $\gamma_C = 0.3$, $\Delta = 1$, $T_C = 1$, and $\gamma_L = \gamma_R = 0.1$. (b) Rectification coefficient $\mathcal{R}$ as a function of $T_L$, reaching a maximum magnitude of about $3.36\%$ at $T_L = 1.7$ and $T_R = 0.3$. (c) Rectification coefficient $\mathcal{R}$ as a function of the collective coupling strength $\gamma_C$, with fixed bias $T_L = 1.7$, $T_R = 0.3$, $\Delta = 1$, and $T_C = 1$. (d) Rectification coefficient $\mathcal{R}$ as a function of the anisotropy parameter $\Delta$, with fixed bias $T_L = 1.7$, $T_R = 0.3$, $\gamma_C = 0.3$, and $T_C = 1$. The positive sign of $\mathcal{R}$ indicates that heat flows more readily from left to right than from right to left. All currents are in units of $\omega^2$.
		}
		\label{Fig_3}
	\end{figure}

	Figure~\ref{Fig_3}(a) and (b) display the currents and the resulting rectification response as a function of $T_L$ under the constraint $T_L + T_R = 2$ (with $\gamma_C = 0.3$ and $\Delta = 1$). The curves exhibit perfect antisymmetry, $\mathcal{R}(T_L, T_R) = -\mathcal{R}(T_R, T_L)$, and vanish at the thermal equilibrium point $T_L=T_R=1$. For finite bias, the rectification remains small but systematic. At the maximum temperature bias considered, $(T_L, T_R) = (1.7, 0.3)$, we find a positive rectification of $\mathcal{R} \approx +3.36\%$, indicating that heat transport is favored from left to right under the chosen convention. As the thermal gradient is reduced, the rectification reaches a local minimum of $\mathcal{R} \approx -1.60\%$ at $T_L = 1.4$ ($T_R=0.6$) before approaching zero at the symmetry point.
	
	To understand how the collective bath controls this asymmetry, we plot $\mathcal{R}$ as a function of the collective coupling strength $\gamma_C$ in Fig.~\ref{Fig_3}(c) under a fixed maximum bias of $T_L = 1.7$ and $T_R = 0.3$, with $\Delta = 1$ and $T_C = 1$. In the absence of collective dissipation ($\gamma_C = 0$), the rectification vanishes, confirming that local coupling to symmetric baths cannot break reciprocity. As $\gamma_C$ increases, $\mathcal{R}$ initially rises to a positive maximum of $\approx +6.55\%$ at $\gamma_C \approx 0.10$, indicating enhanced forward transport for this specific bias. However, further increasing the coupling suppresses the forward current faster than the reverse current, leading to a sign change in rectification at $\gamma_C \approx 0.43$. For larger couplings, the rectification becomes negative, reaching $\mathcal{R} \approx -7.39\%$ at $\gamma_C = 0.8$. We emphasize that these values correspond to a sweep over $\gamma_C$ at fixed bias $(T_L, T_R) = (1.7, 0.3)$, distinct from the bias sweep discussed in Fig.~\ref{Fig_3}(b), which yielded $\mathcal{R} \approx 3.36\%$ at $\gamma_C = 0.3$. The nonmonotonic dependence on $\gamma_C$ indicates the existence of an optimal collective coupling strength for maximizing rectification. Beyond this optimum, further strengthening the collective dissipative channel suppresses the directional current imbalance, leading to a reduction in the rectification coefficient.

	Finally, the role of internal chain interactions is explored in Fig.~\ref{Fig_3}(d), which shows $\mathcal{R}$ as a function of the anisotropy parameter $\Delta$. The rectification remains positive across the entire scanned range of $\Delta \in [-1, 2]$, but exhibits nonmonotonic behavior with a local maximum of $\mathcal{R} \approx 7.25\%$ near $\Delta \approx -0.4$. As $\Delta$ increases beyond the isotropic point, the rectification rises sharply, reaching its maximum value of $\mathcal{R} \approx 20.1\%$ at $\Delta = 2.0$. It is important to note, however, that this large rectification at high anisotropy occurs in a regime where both the forward and reverse heat currents are significantly suppressed. This trade-off is physically significant: the same Ising-like interactions that suppress heat transport also enhance the asymmetry between forward and reverse currents. The practical consequence is that optimizing the rectification coefficient $\mathcal{R}$ alone is insufficient; a useful thermal diode must balance rectification against throughput.
	
	\section{Thermodynamic Cost and Entropy Production}
	\label{sec:entropy_production}
	
	While the preceding sections focused on the magnitude, direction, and controllability of the steady-state heat currents, an equally important aspect of nonequilibrium operation is the irreversible thermodynamic cost required to sustain these currents. This cost is quantified by the total entropy-production rate, denoted by $\dot{\Sigma}$. For a Markovian open quantum system governed by thermodynamically consistent Lindblad dynamics, the Spohn inequality guarantees \cite{spohn1978}
	\begin{equation}
		\dot{\Sigma} \geq 0,
	\end{equation}
	with equality attained in the absence of irreversible thermodynamic driving. In contrast to the von Neumann entropy of the system, whose time derivative vanishes at steady state, $\dot{\Sigma}$ measures the entropy generated in the combined system--reservoir dynamics.
	
	In the present three-terminal configuration, the entropy-production rate is determined by the heat currents exchanged with the left, right, and collective reservoirs. Adopting the convention that \(J_\alpha>0\) denotes heat flowing from reservoir \(\alpha\) into the spin chain, we define
	\begin{equation}
		\dot{\Sigma}
		=
		-\sum_{\alpha\in\{L,R,C\}}
		\frac{J_\alpha}{T_\alpha}
		=
		-\sum_{\alpha\in\{L,R,C\}}
		\beta_\alpha J_\alpha,
		\label{eq:entropy_production}
	\end{equation}
	where \(\beta_\alpha=1/T_\alpha\) is the inverse temperature of reservoir \(\alpha\). The entropy-production rate defined in Eq. (\ref{eq:entropy_production}) provides the thermodynamic measure used here to characterize the irreversibility associated with the steady-state currents generated by the adopted local Lindblad model. Although our system is interacting, we work in the weak-coupling regime where the local master equation provides a valid description of the dissipative dynamics \cite{alicki1979, spohn1978}. Throughout this section, all temperatures are measured in units of the transition frequency \(\omega\), so that \(\dot{\Sigma}\) is expressed in units of \(\omega\) (with \(\hbar = k_B = 1\)). If the opposite current convention is used, the overall sign in Eq.~\eqref{eq:entropy_production} must be reversed accordingly. This quantity directly connects the heat-transport properties discussed in Sec.~\ref{sec:heat_transport} with the irreversibility of the three-terminal device.
	
\begin{figure}[htbp]
		\centering
		\includegraphics[width=0.45\textwidth]{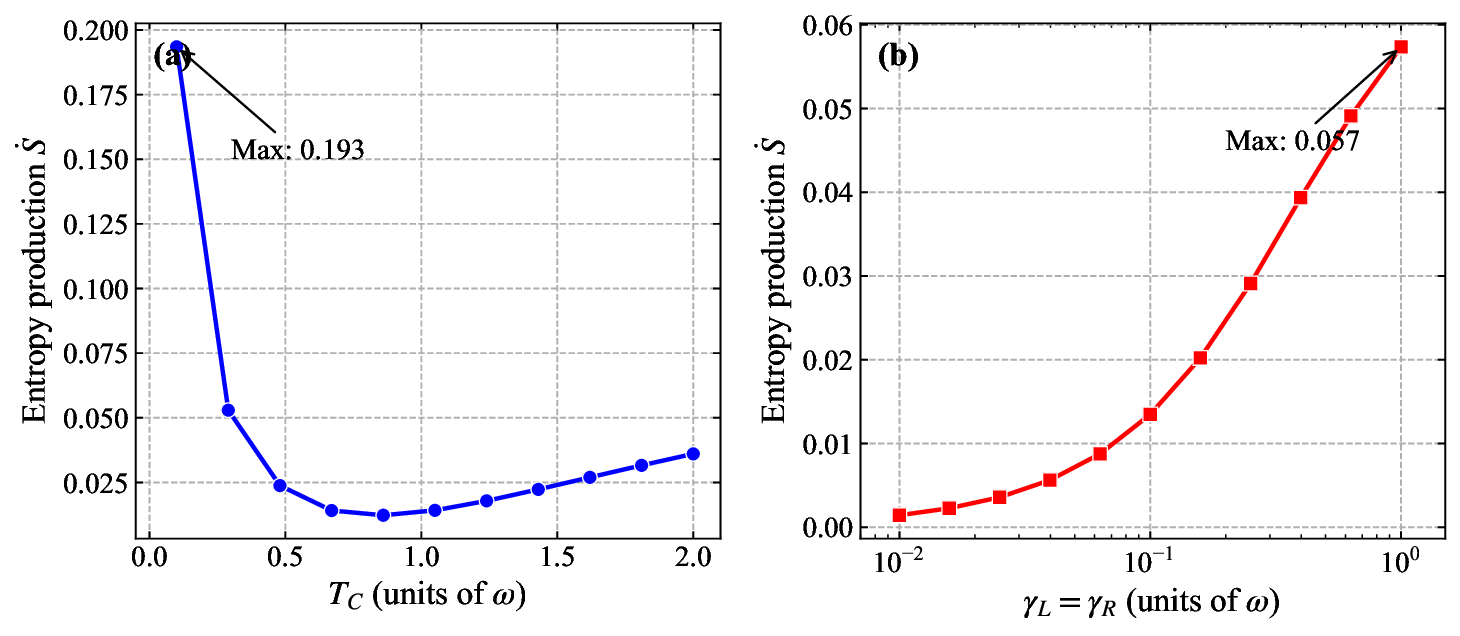}
		\caption{
			Steady-state entropy production rate \(\dot{\Sigma}\) (in units of \(\omega\)) as a function of environmental control parameters. 
			\textbf{(a)} \(\dot{\Sigma}\) versus the collective-reservoir temperature \(T_C\) on a linear scale, with fixed parameters \(\gamma_L=\gamma_R=0.1\omega\), \(T_L=1.5\omega\), \(T_R=0.5\omega\), \(\gamma_C=0.3\omega\), and \(\Delta=1\). 
			\textbf{(b)} \(\dot{\Sigma}\) versus the symmetric local dissipation strengths \(\gamma_L = \gamma_R\) on a logarithmic scale, with fixed parameters \(T_L=1.5\omega\), \(T_R=0.5\omega\), \(T_C=1.0\omega\), \(\gamma_C=0.3\omega\), and \(\Delta=1\). 
		}
		\label{Fig_4}
	\end{figure}

	Figure~\ref{Fig_4} shows the dependence of the steady-state entropy-production rate on the collective-reservoir temperature \(T_C\) and on the symmetric boundary dissipation strength \(\gamma_L=\gamma_R\). These two control parameters affect the thermodynamic cost in qualitatively distinct ways.
	
	Figure~\ref{Fig_4}(a) displays $\dot{\Sigma}$ as a function of \(T_C\). The dependence is strongly non-monotonic. At the lowest collective-reservoir temperature considered, \(T_C=0.1\omega\), the device exhibits its largest entropy-production rate, $\dot{\Sigma}_{\max}\simeq 0.193,$ reflecting the strong thermodynamic affinity produced by the large temperature imbalance among the three terminals. As $T_C$ increases, $\dot{\Sigma}$ decreases rapidly and reaches a minimum around \(T_C\simeq 0.86\omega\). This regime therefore represents a low-irreversibility operating point of the device.
	
	The minimum of $\dot{\Sigma}$ occurs close to the temperature at which the collective heat current \(J_C\) changes sign (approximately \(T_C \simeq 1.0\omega\), as shown in Fig.~\ref{Fig_2}(c)). The proximity of these two features—the entropy minimum and the current reversal—is physically meaningful. When the collective bath is tuned to the regime where it neither strongly absorbs nor emits heat, the redistribution of energy among the three reservoirs is accompanied by reduced global irreversibility. This suggests that the collective bath can be used not only to control the direction of heat flow but also to minimize the thermodynamic cost of operating the device. Near this crossover, the collective reservoir changes its net thermodynamic role, from extracting energy from the chain to injecting energy into it, or conversely according to the adopted current convention. For \(T_C\gtrsim 1.0\omega\), the entropy-production rate increases again, showing that excessive heating of the collective reservoir drives the system away from this low-dissipation regime.
	
	A qualitatively different behavior is observed when varying the symmetric local coupling strength $\gamma_L=\gamma_R$, as shown in Fig.~\ref{Fig_4}(b). The coupling axis is logarithmic and spans the interval from \(10^{-2}\omega\) to \(1.0\omega\). As the boundary couplings become weak, the associated heat currents decrease, leading to a corresponding reduction of the entropy-production rate. As \(\gamma_L=\gamma_R\) increases, the entropy-production rate rises monotonically, reaching $\dot{\Sigma}\simeq 0.057$ at \(\gamma_L=\gamma_R=1.0\omega\).
	
	This monotonic behavior is physically intuitive: stronger local coupling enhances the rate of energy exchange with the boundary reservoirs, which in turn drives the system further from equilibrium and increases irreversibility. The weak-coupling regime minimizes entropy production, but it also suppresses the heat-current magnitude and increases the time required to approach the steady state. Therefore, minimizing $\dot{\Sigma}$ alone does not define an operational optimum; an efficient operating regime must balance low irreversibility against appreciable heat transport and a practically accessible relaxation time.
	
	Taken together, Fig.~\ref{Fig_4} demonstrates that the collective reservoir provides a nontrivial environmental control channel for the thermodynamic cost of the three-qubit XXZ chain. In particular, tuning \(T_C\) permits access to a low-entropy-production regime near \(T_C\simeq 0.86\omega\), whereas increasing the local boundary dissipation monotonically increases the irreversible cost over the investigated range. These results complement the rectification analysis of Sec.~\ref{sec:rectification}: the collective environment can be used not only to alter the heat-current asymmetry, but also to regulate the thermodynamic overhead associated with nonequilibrium transport.
	
	Throughout the parameter ranges considered here, $\dot{\Sigma}$ remains non-negative, consistent with the second-law behavior expected within the thermodynamic framework adopted for the present local Lindblad model. \cite{Hewgill2021}.
	
\section{Conclusion}
\label{sec:conclusion}

In this work, we investigated the steady-state heat transport, thermal rectification, and thermodynamic cost of a minimal three-qubit XXZ spin chain driven by three terminals: two local reservoirs coupled to the boundary qubits and a collective reservoir jointly coupled to the middle and right qubits. Our analysis demonstrates that engineered collective dissipation can serve as an effective thermodynamic control resource, allowing the direction, magnitude, and irreversibility of quantum heat transport to be manipulated without modifying the internal Hamiltonian.

We found that the collective bath actively participates in energy exchange rather than merely providing additional damping. The collective current $J_C$ changes sign as a function of the collective bath temperature $T_C$, signaling a crossover from net energy extraction to net energy injection. The anisotropy parameter $\Delta$ strongly suppresses boundary heat currents, consistent with excitation localization in the Ising-dominated regime.

This asymmetric dissipation enables thermal rectification, reaching $\mathcal{R} \approx 3.36\%$ at maximum bias under moderate anisotropy and up to $\mathcal{R} \approx 20.1\%$ in the strongly anisotropic (Ising) regime, although the latter comes at the cost of severely suppressed heat currents. Interestingly, increasing the collective coupling can even reverse the direction of rectification, demonstrating that the collective dissipative channel provides a tunable means of controlling the transport asymmetry. This trade-off between rectification strength and heat-current magnitude is a central feature of the device.

The entropy-production rate $\dot{\Sigma}$ reveals that the collective bath temperature can be tuned to minimize irreversibility near $T_C \approx 0.86\omega$, which lies close to the $J_C$ sign reversal at $T_C \approx 1.0\omega$. The small offset between these features highlights the role of the boundary currents in determining the global entropy balance. Increasing local boundary dissipation monotonically raises the thermodynamic cost, while weak boundary coupling minimizes entropy production but at the cost of reduced heat currents and slower relaxation.

Our findings connect to recent work on auxiliary-atom-controlled thermal diodes \cite{zhang2025}, where environmental degrees of freedom modulate rectification. Here, we extend this paradigm by showing that the collective bath itself—rather than an auxiliary internal degree of freedom—can serve as the control knob, and we further quantify the associated thermodynamic cost, which was not addressed in previous studies. The entropy production analysis further establishes a direct link between transport and irreversibility, complementing recent studies of thermodynamic cost in quantum thermal devices \cite{ptaszynski2023}.

The proposed architecture is compatible with platforms such as superconducting circuits and trapped ions, where collective dissipation can in principle be engineered through shared electromagnetic environments or common lossy modes \cite{you2011, blatt2012, vanloo2013, shankar2013, wendin2017}. The tunability of collective decay rates and bath temperatures provides realistic routes for experimental verification. Future work could explore the role of non-Markovian effects, time-modulated baths, or extensions to longer spin chains to enhance rectification while reducing entropy production.

\end{document}